\documentclass[aps,twocolumn,showpacs,preprintnumbers,floatfix,nofootinbib]{revtex4-1}
\usepackage{amsmath}    
\usepackage{amssymb}
\usepackage{graphicx}
\usepackage{graphics}
\usepackage{epsfig}
\usepackage{verbatim}   
\usepackage[subfigure]{graphfig}
\usepackage{epsfig}
\usepackage{epstopdf}
\usepackage{dcolumn}
\usepackage{color}
\usepackage{tabularx}

 \def\bea{\begin{eqnarray}}
 \def\eea{\end{eqnarray}}

\newcommand{\nn}{\nonumber}

\definecolor{green}{rgb}{0,.5,0}

\begin{document}


\title{\vspace{1.0in} {\bf Glue spin and helicity in proton from lattice QCD}}

\author{Yi-Bo Yang$^{1}$, Raza Sabbir Sufian$^{1}$, Andrei Alexandru$^{2}$,  Terrence Draper$^{1}$,  Michael J. Glatzmaier$^{1}$, 
Keh-Fei Liu$^{1}$ and Yong Zhao$^{3,4}$
\vspace*{-0.5cm}
\begin{center}
\large{
\vspace*{0.4cm}
\includegraphics[scale=0.20]{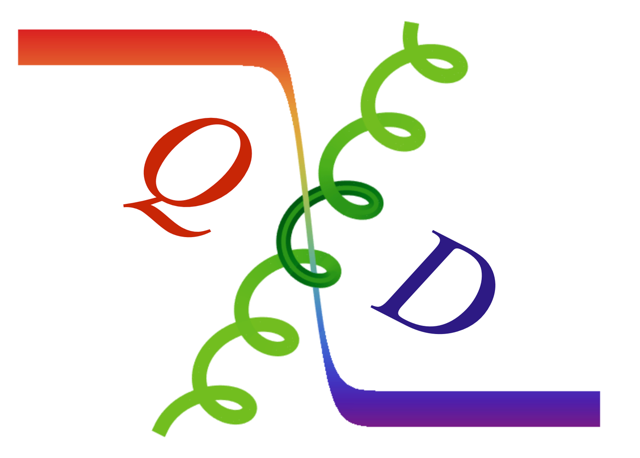}\\
\vspace*{0.4cm}
($\chi$QCD Collaboration)
\vspace*{0.2cm}
}
\end{center}
}
\affiliation{
$^{1}$\mbox{Department of Physics and Astronomy, University of Kentucky, Lexington, KY 40506, USA}\\
$^{2}$\mbox{Department of Physics, The George Washington University, Washington, DC 20052, USA}\\
$^{3}$\mbox{Maryland Center for Fundamental Physics, University of Maryland, College Park, Maryland 20742, USA}\\
$^{4}$\mbox{Nuclear Science Division, Lawrence Berkeley National Laboratory, Berkeley, CA, 94720, USA}\\
}

\begin{abstract}
We report the first lattice QCD calculation of the glue spin in the nucleon. The lattice calculation is carried out with valence overlap fermions on 2+1 flavor DWF gauge configurations on four lattice spacings and four volumes including an ensemble with physical values for the quark masses. The glue spin $S_{G}$ in the Coulomb gauge in the $\overline{\text{MS}}$ scheme is obtained with the 1-loop perturbative matching. We find the results fairly insensitive to lattice spacing and quark masses. We also find that the proton momentum dependence of $S_G$ in the range $0\leq |\vec{p}| < 1.5$ GeV is very mild, and we determine it in the large momentum limit to be $S_{G}=0.251(47)(16)$ at the physical pion mass in the $\overline{\text{MS}}$ scheme at $\mu^2=10$ GeV$^2$. If the matching procedure in large momentum effective theory is neglected, $S_G$ is equal to the glue helicity measured in high-energy scattering experiments.
\end{abstract}


\maketitle


{\it Introduction:} Deep-inelastic scattering experiments reveal that contrary to the naive quark model, the quark spin contribution to the proton spin is quite small, about 30$\%$~\cite{Accardi:2012qut,deFlorian:2009vb,Aidala:2012mv}. In an effort to search for the missing proton spin, recent analyses~\cite{deFlorian:2014yva,Nocera:2014gqa} of the high-statistics 2009 STAR~\cite{Djawotho:2013pga} and PHENIX~\cite{Adare:2014hsq} {experiments at RHIC} showed evidence of non-zero glue helicity $\Delta G$ in the proton.  For $Q^2=10$ GeV$^2$, the glue helicity distribution $\Delta g(x,Q^2)$ is found to be positive and away from zero in the momentum fraction region $x<0.05$. However, the results are limited by very large uncertainty in this region. 

The recent COMPASS analysis explored $\Delta g(x)$ from the scaling violation of $\Delta q(x)$, and the highly distinct solutions of $\Delta g(x)$ can be obtained with different parameterizations of $\Delta q(x)$~\cite{Adolph:2015saz}. Therefore, it hints that if a high precision $\Delta g(x)$ can be obtained directly, it will benefit our understanding of the parameterizations of $\Delta q(x)$ and provide more information about the role of quark spin in the proton.

Given the importance of $\Delta g(x)$ to explain the origin of the proton spin, and the fact that significant efforts are devoted to its precise experimental determination, a theoretical understanding and calculation of $\Delta G$ is highly desired. $\Delta G$ is defined as the first moment of the glue helicity distribution $\Delta g(x)$~\cite{Manohar:1991ux},
\begin{eqnarray} \label{eq3}
\Delta G &=& \int dx \frac{i}{2xP^+}\int \frac{d\xi^-}{2\pi}e^{-ixP^+\xi^-} \nonumber\\
&&\quad \langle PS | F^{+\alpha}_a (\xi^-)\mathcal{L}^{ab}(\xi^-,0)\tilde{F}^+_{\alpha,b}(0) | PS\rangle\ ,
\end{eqnarray}
where the light front coordinates are $\xi^{\pm}=(\xi^0\pm\xi^3)/\sqrt{2}$. The proton plane wave state is written as $|PS\rangle$, with momentum $P^\mu=(P,0,0,P)$ and polarization $S$. The light-cone gauge-link $\mathcal{L}(\xi^-,0)=P \textrm{exp}[-i g\int^{\xi-}_0 A^+(\eta^-,0_{\perp})d \eta^-]$ is defined in the adjoint representation.
 It connects the gauge field tensor and its dual, $\tilde{F}^{\alpha\beta}=\frac{1}{2}\epsilon^{\alpha\beta\mu\nu}F_{\mu\nu}$, to construct a gauge invariant operator. 
After integrating over $x$, one can define the gauge-invariant gluon helicity operator in a non-local form~\cite{Hatta:2011zs,Ji:2013fga},
\begin{equation}\label{eq4}
\tilde{S}_g = \Bigg[ \vec{E}^a(0)\times (\vec{A}^a(0)-\frac{1}{\nabla^+} (\vec{\nabla}A^{+,b})\mathcal{L}^{ba}(\xi^-,0))\Bigg]^z
\end{equation}
where $\nabla^+=\partial/\partial \xi^-$. It is the gauge-invariant extension (GIE) of the operator $\vec{E}\times \vec{A}$ in the light-cone gauge $A^{+}=0$, but
one cannot evaluate this expression on the lattice directly due to its real-time dependence. 

On the other hand, $\tilde{S}_g$ is equal to the infinite momentum frame (IMF) limit of a universality class of operators~\cite{Hatta:2013gta} whose matrix elements can be matched to $\Delta G$ through a factorization formula in large momentum effective theory (LaMET)~\cite{Ji:2014gla,Ji:2014lra}. 
The gluon spin operator proposed in Ref.~\cite{Chen:2008ag, Chen:2009mr} with the non-abelian transverse condition belongs to this universality class
and has been proven to be equivalent to the GIE of $\vec{E}\times \vec{A}$ in the Coulomb gauge $\vec{\partial}\cdot\vec{A}=0$~\cite{Lorce:2012rr,Zhao:2015kca},
\begin{eqnarray}\label{ExA}
\vec{S}_g =  2 \int d^3x\ \textrm{Tr} (\vec{E}_c\times \vec{A}_c),
\end{eqnarray}
where the factor 2 is from the normalization of the $SU(3)$ group generators and $\vec{E}_c$ and $\vec{A}_c$ are the chromoelectric field and gauge potential in the Coulomb gauge with their lattice versions to be addressed in the following.

$\vec{S}_g$ is not Lorentz covariant and has nontrivial frame dependence~\cite{Ji:2013fga}. It is shown in Ref.~\cite{Hatta:2013gta} that when boosted to the IMF, the Coulomb gauge fixing condition (as well as the temporal condition $A^0=0$)~\cite{Hatta:2013gta} become $A^+=0$, and then the longitudinal component of $\vec{S}_g$ in either gauge is equivalent to the glue helicity operator $\tilde{S}_g$ with a proper matching to cancel the intrinsic frame dependence of $\vec{S}_g$.
 On the lattice, the Coulomb condition can be obtained numerically~\cite{Schrock:2012fj} and the glue spin operator $\vec{S}_g$ in the Coulomb gauge can be calculated without numerical difficulty.

The major task of this work is calculating the matrix element of $\vec{S}_{g}$ in the proton, which will be indicated as $S_{G}$, in the rest and moving frames. The results are then renormalized at 1-loop order in lattice perturbation theory and matched to the $\overline{\textrm{MS}}$ scheme at $\mu^2$=10 GeV$^2$, to investigate their frame dependence and address the matching to the helicity.

{\it Numerical details:} A preliminary attempt ~\cite{Sufian:2014jma} to calculate $S_{G}$ was carried out on $2+1$ flavor dynamical domain-wall configurations on a $24^3\times 64$ lattice (24I) with the sea pion mass at $330$ MeV and on a $32^3\times 64$ lattice with sea pion mass at 300 MeV~\cite{Yang:2016nfc}. In this work, we improve the statistics on the ensembles mentioned above and carry out the calculation on another three ensembles with different lattice spacings, volumes, and sea quark masses to check the corrections to the glue spin from various systematic uncertainties. We use the 2-2-2 smeared stochastic grid source on all the ensembles (except 48I where the 4-4-4 smeared stochastic grid source is used), and apply the low-mode substitution~\cite{Li:2010pw,Gong:2013vja} to make the signal-to-noise ratio close to that with 8 (64 on the 48I ensemble) independent smeared point sources. Furthermore, we loop over all the time slices for the two-point functions of the nucleon to increase statistics. The statistics used for this grid source measurements is roughly equivalent to evaluating a large number of quasi-independent smeared point source measurements ranging from 103,936 on the 24I lattice to 497,664 on the 48I lattice. The parameters of the ensembles used in this work are listed in Table~\ref{table:r0} and more details of the simulation setups can be found in the supplemental materials~\cite{sm_glue_spin}.

\begin{table}
\begin{center}
\caption{\label{table:r0} The parameters for the RBC/UKQCD configurations~\cite{Blum:2014tka}. $m_{\pi}^{(s)}$ is the pion mass of the light sea quark in the 2+1 flavor configuration, and $N_{cfg}$ is the number of configurations used in the simulation.}
\begin{tabular}{c|cccc}
\hline
Symbol & $L^3\times T$ & $a(\textrm{fm})$ & $m^{(s)}_{\pi}$\textrm{(MeV)}  & $N_{cfg}$ \\
 \hline
32ID & $32^3\times 64$ & 0.1431(7) &  170 & 200 \\
 \hline
48I &$48^3\times 96$ & 0.1141(2)  &  140 &\  81 \\
 \hline
24I & $24^3\times 64$ & 0.1105(3) &  330 & 203 \\
 \hline
32I & $32^3\times 64$ & 0.0828(3) &  300 & 309 \\
 \hline
32If  &$32^3\times 64$ & 0.0627(3) &  370 & 238 \\
  \hline
\end{tabular} 
\end{center}
\end{table}

The Coulomb gauge fixing condition used here is enforced by requiring that the spatial sum of the backward difference of the HYP-smeared gauge links \cite{Hasenfratz:2001hp} to be zero,
 \begin{eqnarray}
 \sum_{\mu=x,y,z} \Big[ U^c_{\mu}(x)-U^c_{\mu}(x-a\hat{\mu})\Big]=0,
 \end{eqnarray}
where $U^c_\mu(x)$ is the Coulomb gauge fixed Wilson link from $x+a\hat{\mu}$ to $x$. The gauge fixed potential $A_c$ is defined by 
 \begin{eqnarray}
\!\!\!\!\!A_{c,\mu}=\Big[\frac{U^c_\mu(x)-U^{c\dagger}_\mu(x)+U^c_\mu(x-a\hat{\mu})-U^{c\dagger}_\mu(x-a\hat{\mu})}{4iag}\Big]_{\textrm{traceless}}
 \end{eqnarray}
 with $g$ as the bare coupling constant,
and the chromoelectric field used in this work is given by the clover definition
\begin{eqnarray}
F_{\mu\nu}^c &=& \frac{i}{8a^2g} (\mathcal{P}_{\mu,\nu}-\mathcal{P}_{\nu,\mu}+\mathcal{P}_{\nu,-\mu}-\mathcal{P}_{-\mu,\nu} \nn \\
&& + \mathcal{P}_{-\mu,-\nu} -\mathcal{P}_{-\nu,-\mu} + \mathcal{P}_{-\nu,\mu} -\mathcal{P}_{\mu,-\nu}),
\end{eqnarray}
 where \mbox{$\mathcal{P}_{\mu,\nu} = U^c_\mu(x)U^c_\nu(x+a\hat{\mu})U^{c\dagger}_\mu(x+a\hat{\nu})U^{c\dagger}_\nu(x)$}.  

 \begin{figure}[htb]
\centering
\includegraphics[scale=0.25]{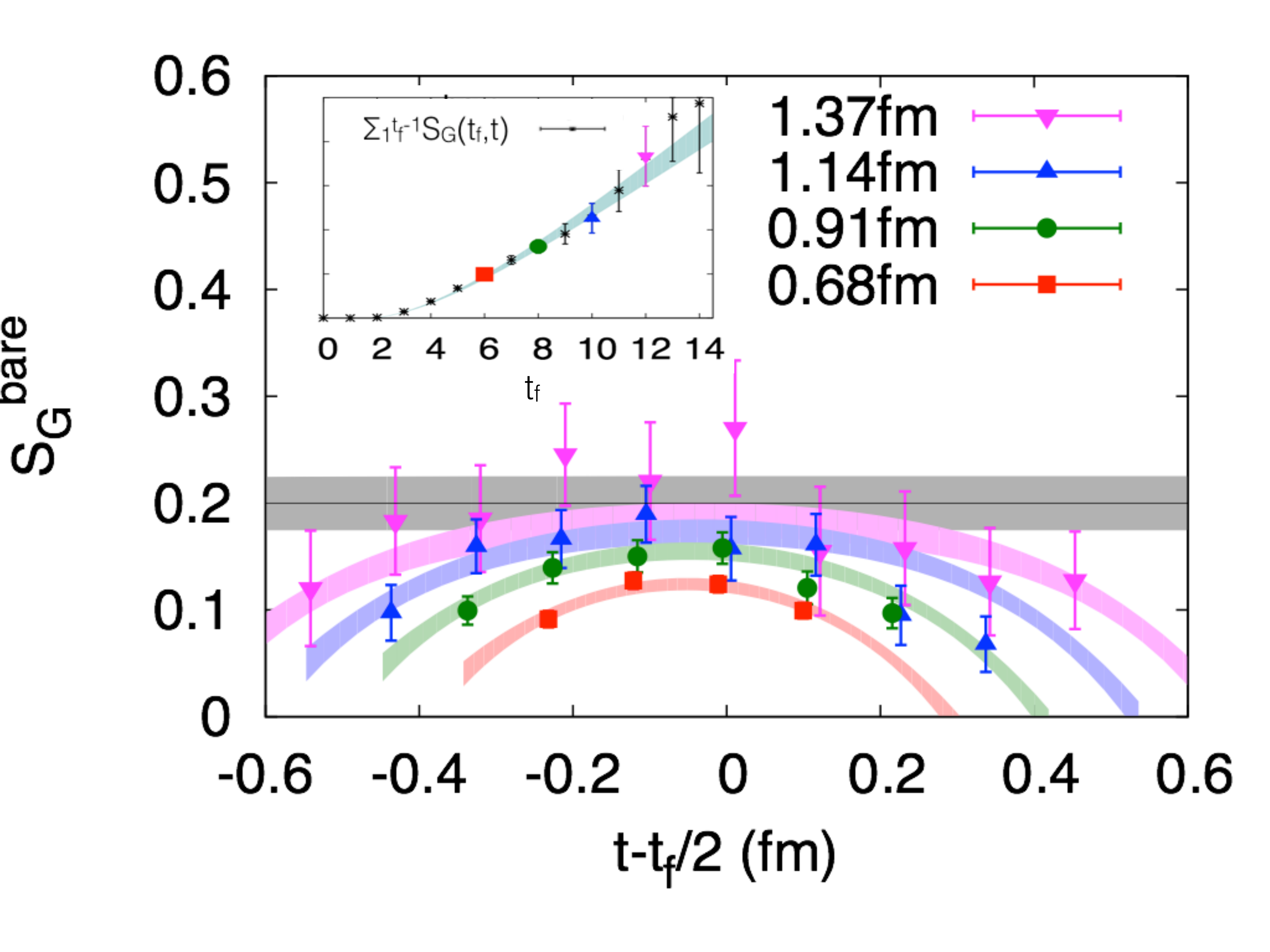}
\vspace*{-0.5cm}
 \caption{\small The ratio $R(t_f,t)$ as a function of the source-sink separation $t_f$ and the current time slice $t$, for the bare glue spin matrix element in the proton, $S_{G}^{bare}$, is plotted at the unitary point on the 24I ensemble. The gray band shows the result extrapolated to infinite separation, which corresponds to the prediction of $S_{G}$. The excited-state contamination is small when the source-sink separation is larger than 1~fm. The sub-panel in the plot shows that the prediction of $\sum_t R(t_f,t)$ from the two-state fit (the band) in Eq.~(\ref{eq:two_term}) agrees well with the data points.}
\label{fig:two_term}
\end{figure}

In order to extract $S_{G}$, we compute the ratio of the disconnected three-point function with the gluon operator insertion to the nucleon propagator  with the source and sink of the nucleon located at $0$ and $t_f$, respectively. 
The glue spin operator is inserted at the time slice $t$ which is between $0$ and $t_f$. 
Then the ratio in a moving frame $\vec{p}=(0,0,p_3)$ along the $z$-direction is
\begin{eqnarray}\label{eq:R_def}
R(t_f,t) = \frac{\langle 0|\Gamma^m_{3}\int d^3y\ e^{-ip_3y_3}\chi(\vec{y},t_f)S_{g}^{3}(t)\bar{\chi}(\vec{0},0)|0 \rangle}{\langle 0|\Gamma^e\int d^3y\ e^{-ip_3y_3}\chi(\vec{y},t_f)\bar{\chi}(\vec{0},0)|0 \rangle}, \nn \\
\end{eqnarray}
where $\chi$ is the nucleon interpolation field and $\Gamma^e$ and $\Gamma^m_{3}$ are the unpolarized projection operator of the proton and the polarized one along the $z$-direction, respectively. 
When $t_f$ is large enough, $R(t_f,t)$ is equal to 
the proton matrix element of the longitudinal glue spin operator, $S_{G}$, plus $t$-dependent corrections,
\begin{eqnarray}\label{eq:two_term}
R(t_f,t) ={S_{G}}+C_1 e^{-\Delta E(t_f-t)}+ C_2 e^{-\Delta E t}+ C_3 e^{-\Delta Et_f},\nonumber\\
\end{eqnarray}
where $\Delta E$ is the energy difference between the first excited state and the ground state and $C_{1,2,3}$ are the spectral weights involving the excited state.

We plot the ratio $R(t_f,t)$ for the unitary point on the 24I ensemble, as a function of $t-t_f/2$ for several $t_f$ in Fig.~\ref{fig:two_term}. The curves predicted by the fit agree with the data, and the $\chi^2$/d.o.f is smaller than 1.4 for all the other quark masses on five ensembles. From the fit, we see that the excited-state contamination is small when the source-sink separation is larger than 1 fm. The final prediction of $S_{G}$ (the gray band) is consistent with the blue and purple data points  at $t\sim t_f/2$. Similar plots for the other ensembles can be found in the supplemental materials~\cite{sm_glue_spin}.

\begin{figure}[htb]
\centering
\includegraphics[scale=0.7]{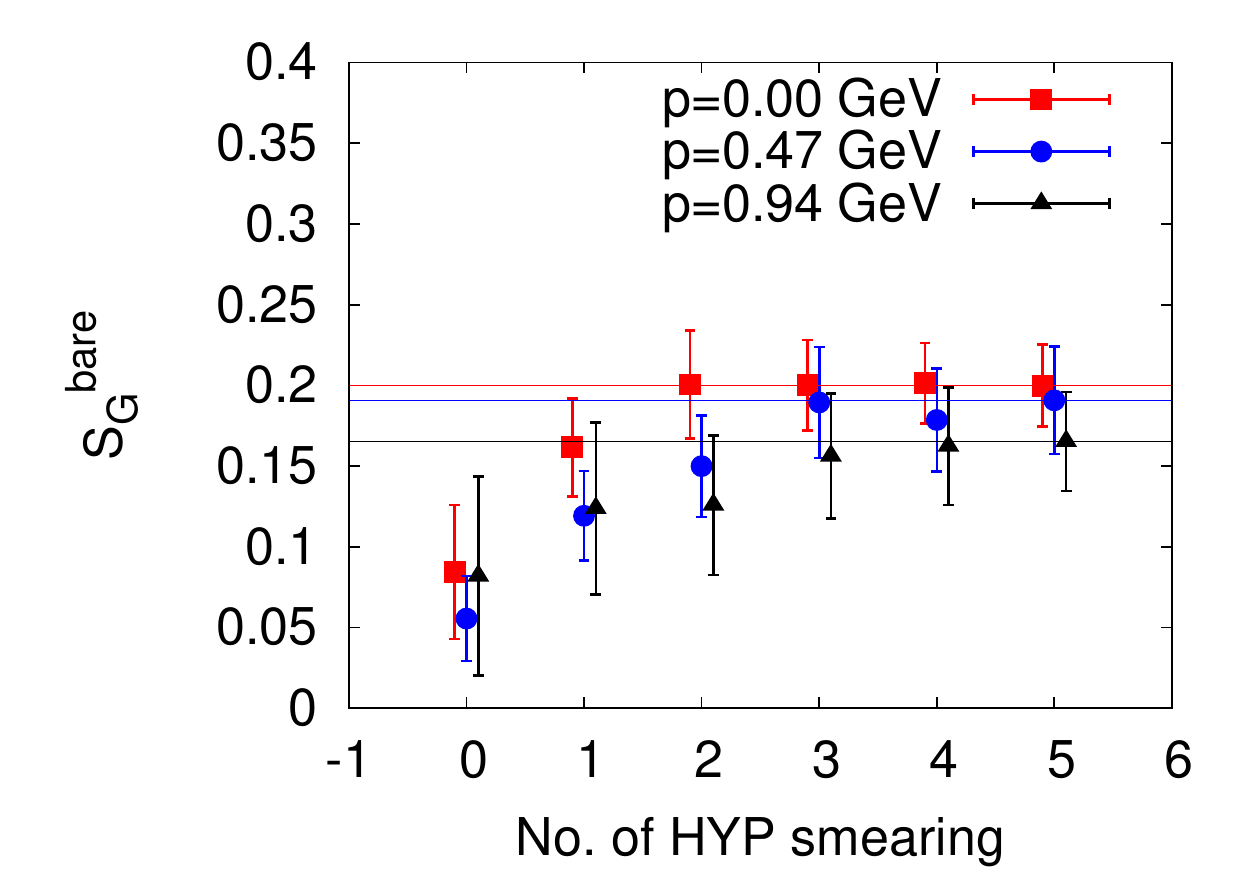}
\vspace*{-0.5cm}
 \caption{\small The HYP smearing steps dependence of the bare glue spin $S_{G}$ at the unitary point on the ensemble 24I, for $p=0$ (red squares), $p=0.47$ GeV (blue dots) and $p=0.94$ GeV (black triangles). The values of $S_{G}$ are unchanged after two or three steps of smearing.}
\label{fig:dep}
\end{figure}

 It is observed that the central values of the glue spin matrix elements as a function of HYP smearing steps are unchanged after two or three steps of smearing, as shown in Fig.~\ref{fig:dep} (for the case of the unitary point on the ensemble 24I), while the signal to noise ratio (SNR) can be improved when more HYP smearing steps are applied. In this work, 5 steps of HYP smearing are used for the glue spin operator on each ensemble, and the nucleon two-point correlators with the source located on all the time slices are generated to increase the SNR. Since the tadpole improvement factor is $1/u_0^5 \sim$ 2 for the $S_{g}$ operator without any HYP smearing, the enlargement of the result after the HYP smearing is understandable. Note that the HYP smearing here just affects the glue spin operator but the gauge action is unchanged since no reweighting is applied on configuration averages.

{\it Results:}  The renormalized matrix element $S_{G}$ including mixing from the quark spin is~\cite{yang:2016lpt_exa}
\begin{eqnarray}\label{eq:mixing}
S^{\overline{\text{MS}}}_{G}&=&
\ \ \left\{1- \frac{g^2}{16\pi^2}\left[N_f\left(\frac{2}{3}\textrm{log}(\mu^2a^2)+1.27\right)\right.\right.\nonumber\\
&&\left.\left.\quad\quad-C_A\left(\frac{4}{3}\textrm{log}(\mu^2a^2)
+f_{gg}(g^2)\right)\right]\right\}S^{L}_{G}\nonumber\\
&&+\frac{g^2C_F}{16\pi^2}\left(\frac{5}{3}\textrm{log}(\mu^2a^2)+7.31\right) \Delta\Sigma^{L}\nonumber\\ &&+O(g^4)\ ,
\end{eqnarray}
where the superscripts $\overline{\text{MS}}$ and $L$ indicate the quantities under the $\overline{\text{MS}}$ scheme and that under lattice regularization.  We applied the Cactus improvement \cite{Constantinou:2006hz} to re-sum the major tadpole contributions to get a better convergence in the 1-loop correction of the glue spin. Then the re-summed finite piece $f_{gg}(g^2)$ depends on the bare coupling $g^2$ weakly and is in the range of 1.7--2.4 for the values of $g^2$ we used in this work. The details are addressed in Ref.~\cite{yang:2016lpt_exa}. Since the value of the mixing term involving $\Delta \Sigma^L$ in Eq.~(\ref{eq:mixing}) is at the same order of the present statistical error of $S_G^L$, the uncertainty due to $\Delta \Sigma^L$ for the gauge ensembles considered here will be even smaller. Therefore, we approximate the quark spin $ \Delta\Sigma^{L}$ by the experimental value $\Delta\Sigma^{\overline{\text{MS}}}$, which is $\sim$30\% of the total proton spin from the global analysis of deep inelastic scattering data~\cite{Accardi:2012qut,deFlorian:2009vb,Aidala:2012mv}. 

\begin{figure}[h]
\centering
\includegraphics[scale=0.7]{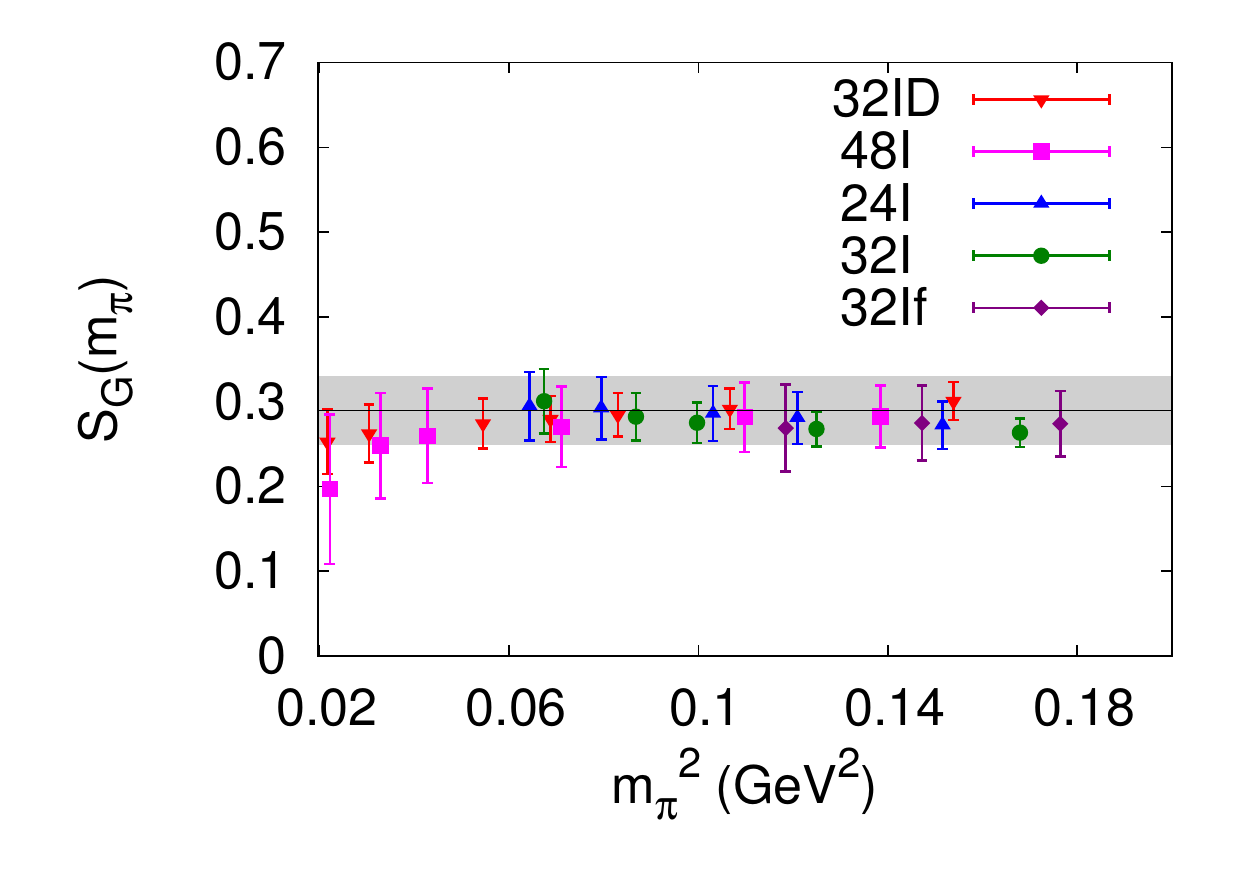}
\vspace*{-0.5cm}
 \caption{\small The valence pion mass dependence of $S_{G}$ at $\mu^2=10$ GeV$^2$, in the rest frame of the proton. These dependencies are fairly mild and can be well described with a linear fit. The gray band shows the result based on the global fit with the empirical form in Eq.~(\ref{eq:global_fit}).}
\label{fig:chiral}
\end{figure}

\begin{figure}[h]
\centering
\includegraphics[scale=0.7]{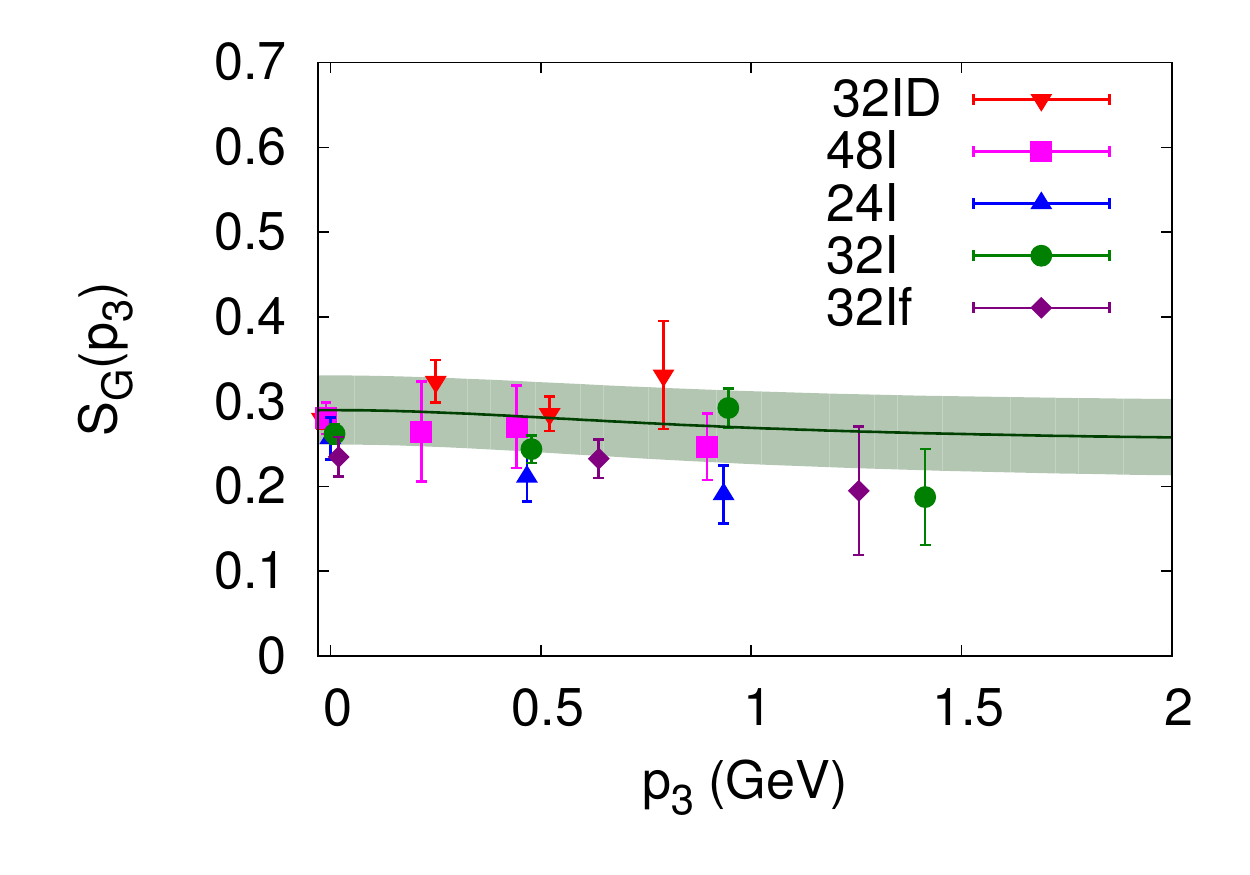}
\vspace*{-0.5cm}
 \caption{\small The results extrapolated to the physical pion mass as a function of the absolute value of $\vec{p}=(0,0,p_3)$, on all the five ensembles. All the results have been converted to $\overline{\text{MS}}$ at $\mu^2=10$ GeV$^2$. The data on several ensembles are shifted horizontally to enhance the legibility.  The green band shows the frame dependence of the global fit (with the empirical form in Eq.~(\ref{eq:global_fit})) of the results.}
\label{fig:final}
\end{figure}

After matching to the values under the $\overline{\textrm{MS}}$ scheme at $\mu^2=10$ GeV$^2$, we find that the valence quark mass dependence is mild regardless of the proton momentum. In Fig.~\ref{fig:chiral}, we show results in the rest frame for various valence quark masses on all the five ensembles, with five pairs of volumes and lattice spacings. Their dependence is also mild. To obtain $S_{G}$ in a relatively large momentum frame, we calculate $S_{G}$ for all the momenta smaller than $\pi/(4a)$ on all the five ensembles. To show the frame dependence, we extrapolate $S_{G}$ on all the ensembles in different momentum frames to the physical value of the valence pion mass, as shown in Fig.~\ref{fig:final}. Some points are not shown in the figure if their uncertainties are larger than the signal. 

The glue helicity in the proton, $\Delta G$, corresponds to the glue longitudinal spin component $S_{G}$ in the IMF. The large-momentum effective field theory (LaEMT)~\cite{Ji:2014lra}, shows a large finite correction at the 1-loop level,
\begin{eqnarray}\label{eq:matching}
S_{G}(|\vec{p}|,\mu)&=&\left[1+\frac{g^2C_A}{16\pi^2}\left(\frac{7}{3}\textrm{log}\frac{(\vec{p})^2}{\mu^2}-10.2098\right) \right]\Delta G(\mu)\nonumber\\
&&+\frac{g^2C_F}{16\pi^2}\left(\frac{4}{3}\textrm{log}\frac{(\vec{p})^2}{\mu^2}-5.2627\right) \Delta \Sigma(\mu) \nonumber\\
&&+O(g^4)+O(\frac{1}{(\vec{p})^2})\ .
\end{eqnarray}

At $\mu^2=10$ GeV$^2$ and $|\vec{p}|=1.5$ GeV the factor before $\Delta G$ is 0.22, which is much smaller than unity and indicates a convergence problem for the perturbative series even after one re-sums the large logarithms. (The factor is 0.80 if the finite piece 10.2098 is removed.) On the other hand, the largest momentum we have on the lattice with acceptable signal is comparable to the proton mass, so the power corrections in Eq.~(\ref{eq:matching}) cannot be neglected and one cannot simply apply this matching condition. Nevertheless, the mild dependence of $S_{G}$ on the proton momentum as in Fig.~\ref{fig:final} leads us to suggest that it could be a small effect to match to the IMF, i.e., $S_{G}\approx \Delta G+O(1/|\vec{p}|^2)$.

Therefore, we neglect the 1-loop LaMET matching and use the following empirical form to fit our data,
\begin{eqnarray}\label{eq:global_fit}
\!\!\!\!S_{G}(|\vec{p}|)&=&S_{G}(\infty)+\frac{C_1}{M^2+|\vec{p}|^2}+C_2(m_{\pi,vv}^2-m^2_{\pi,\textrm{phys}})\nonumber\\
&&+C_3(m_{\pi,ss}^2-m^2_{\pi,\textrm{phys}})+C_4 a^2,
\end{eqnarray}
where $m_{\pi,\textrm{phys}}$=0.139 GeV and $M=0.939$ GeV are the physical pion and proton mass respectively and $m_{\pi,vv/ss}$ are the valence and sea pion masses respectively. The $1/|\vec{p}|^2$ correction in Eq.~(\ref{eq:matching}) is replaced by $1/(M^2+|\vec{p}|^2)$ to include all the data in the fitting.
Since all the coefficients other than $S_{G}(\infty)$ are small, the cross terms and the higher-order terms are ignored. The overall $\chi^2$/d.o.f. is 1.21 with 110 degrees of freedom. In Fig.~\ref{fig:final}, the band of the global fit with the empirical form in Eq.~(\ref{eq:global_fit}) shows that the frame dependence is mild and the central value is changed by less than 10\% from its value in the rest frame to that at $|\vec{p}|\sim$ 1.5 GeV; the change is smaller than the statistical uncertainty.

Since the Coulomb gauge fixing on the lattice has a built-in $O(a)$ correction, we repeated the fit with a linear term in $a$. The central value is changed by about 1\%, while the uncertainty is larger. We take the variance of the central values from two fits as an estimate of this uncertainty. Similarly, the uncertainty from the volume dependence $e^{-m_{\pi_{vv}}L}$ is estimated in the same way and added to the systematic uncertainties in quadrature. In addition, the value of the quark spin $\Delta\Sigma$ is varied by 20\% to cover the value $\sim$0.30~\cite{Accardi:2012qut} and that from Ref.~\cite{Aidala:2012mv}. The final result is $S_{G}(\infty,\mu^2 = 10\ \text{GeV}^2)$ = 0.251(47)(16) with two errors from the statistical and systematic uncertainties.

{\it Summary and outlook:} In this work, we calculated the glue spin in the proton for the first time based on
$\vec{E}\times \vec{A}$ in the Coulomb gauge~\cite{Chen:2008ag, Chen:2009mr}, with various quark masses, lattice spacings, volumes, and proton momenta. The results show mild dependencies on these quantities. After 1-loop perturbative matching from the lattice theory to the continuum and neglecting the matching effect between the glue spin and helicity, we conclude that the gluon helicity $\Delta G(\mu^2 = 10\ \text{GeV}^2) \approx S_{G}(\infty,\mu^2 = 10\ \text{GeV}^2) = 0.251(47)(16)$, which is 50(9)(3)\% of the total proton spin. The Cactus improvement~\cite{Constantinou:2006hz} we used in Eq.~(\ref{eq:mixing}) indicates that uncertainties can be considerable in perturbative QCD, and its reliability should be checked with non-perturbative renormalization in the future. 

On the LaMET side, the convergence problem warrants the matching condition to be calculated at the 2-loop level or higher. On the other hand, if the glue spin in the temporal gauge can be calculated on the lattice, then its LaMET matching in Eq.~(\ref{eq:matching}) can be avoided at the 1-loop level~\cite{Hatta:2013gta}. This possibility is worthy of further investigation~\cite{yang:2016lpt_exa}.
\\\\

{\bf ACKNOWLEDGMENTS}

We thank X. D. Ji and F. Yuan for useful comments and the RBC and UKQCD collaborations for providing us their DWF gauge configurations. This work is supported in part by the U.S. DOE Grant No.\ DE-SC0013065. A.A. is supported in part by the National Science Foundation CAREER grant PHY- 1151648 and by U.S. DOE Grant No. DE-FG02-95ER40907. Y. Z. is supported in part by the U.S. Department of Energy Office of Science, Office of Nuclear Physics under Award Number DE-FG02-93ER-40762 and DE-AC02-05CH11231. \mbox{Y. Y.} also thanks the Institute of High Energy Physics, Chinese Academy of Science for its partial support and hospitality. This material is based upon work supported by the U.S. Department of Energy, Office of Science, Office of Nuclear Physics, within the framework of the TMD Topical Collaboration. This research used resources of the Oak Ridge Leadership Computing Facility at the Oak Ridge National Laboratory, which is supported by the Office of Science of the U.S. Department of Energy under Contract No. DE-AC05-00OR22725. This work also used the Extreme Science
and Engineering Discovery Environment (XSEDE),
which is supported by National Science Foundation grant
number ACI-1053575.

\begin{widetext}
\section*{Supplemental materials}

\subsection{The simulation setup}

\quad For the nucleon propagators, a regular grid with 2 smeared sources in each spatial direction for the 24I, 32I, 32ID and 32If lattices (4 for the 48I lattice since it has a much large volume) are placed
on each of 2 time slices for the 24I, 32I and  32ID lattices (3 for the 48I lattice and 1 for the 32If lattice due to the different sizes in the time direction).
We loop over all the time slices for the nucleon source. The position of the grid is randomly shifted on each time slice.

On the 32ID and 32If ensembles, we use the smeared sink to reduce the excited-states contamination from the sink side. The setup of the nucleon propagators simulation are summarized in Table.~\ref{table:setup}

\begin{table}[htbp]
\begin{center}
\caption{\label{table:setup} The source/sink setup on the ensembles: $N_{grid}$ is the pattern of the smeared points on a grid source with noises. $L^{sm}_{src}$ and $L^{sm}_{sink}$ are the effective smearing size at the source and sink (we didn't apply the sink smearing on 24I, 32I and 48I so $L^{sm}_{sink}$=0 in those cases).}
\begin{tabular}{cccccccc}
Ensemble & 24I & 32I &48I & 32ID & 32If\\
Volume (fm$^4$) & $2.6^3\times 7.1$ & $2.6^3\times 5.4$ & $5.5^3\times 11.0$ & $4.4^3\times 8.8$ & $2.0^3\times 4.0$ \\
\hline
$N_{grid}$ & $2^3\times 2$ & $2^3\times 2$ &$2^3\times 2$ &$4^3\times 3$ &$2^3\times 1$\\
$L^{sm}_{src}$(fm) & 0.55 & 0.56 & 0.56 & 0.50 & 0.48 \\
$L^{sm}_{sink}$(fm) & 0 & 0 & 0 & 0.50 & 0.48 \\
\hline
\end{tabular}
\end{center}
\end{table}

\subsection{The fit of the ratio $R(t_f,t)$}

The functional form of the summed ratio
\bea\label{eq:two_state}
SR(t_f)\equiv\sum_{t=1}^{t_f-1}R(t_f,t)=(t_f-1) S_G +\frac{e^{-\Delta E}-e^{-\Delta E t_f}}{1-e^{-\Delta E}}(C_1+C_2)+ (t_f-1) e^{-\Delta E t_f} C_3
\eea
can be obtained from that of the ratio 
\bea\label{eq:two_state2}
R(t_f,t)=S_G + e^{-\Delta E (t_f-t)} C_1 + e^{-\Delta E t} C_2 +e^{-\Delta E t_f} C_3
\eea
where $\Delta E$ is the energy difference between the first excited-state and the ground state, and $C_{1,2,3}$ are the spectral weights involving the excited-states.

The standard summation method~\cite{Gong:2013vja} approximates $SR(t_f)$ as
\bea
SR(t_f)\simeq t_f S_G + C_1'
\eea
where  all the terms proportional to $e^{-\Delta E t_f}$ in Eq.~(\ref{eq:two_state}) are dropped. This introduces a systematic uncertainty when $\Delta E$ is not very large. At the same time, the classic plateau method cannot distinguish the $e^{-\Delta E t_f} C_3$ term in Eq.~(\ref{eq:two_state2}) from the matrix element $S_G$ we want, except when the calculations are repeated with several values of  $t_f$.  The two-state method can be more reliable than the summation method since it can make use all the information of $R(t_f,t)$, not just their sum over $t$. 

As a cross check, we plot the summed ratio together with the original plot of the ratios for the case of $m_{\pi}\sim 330$ MeV on all the ensembles in Figs.~\ref{fig:24I}-\ref{fig:32If}. In the right panels of these figures, the four colored points correspond to the sum of the ratios at four different $t_f$ which are shown in the left panels.  The bands in the right panels show the prediction of $SR(t_f)$ based on the two-state fits in Eq.~\ref{eq:two_state2}. The data points agree with the gray band very well at larger $t_f$. 

For the cases on the 32ID and 32If ensembles in which the sink smearing is applied (Fig.~\ref{fig:32ID} and Fig.~\ref{fig:32If}), the curvatures as a function of $t-t_f/2$ are symmetric within uncertainty on both sides of $t=t_f/2$, while those on the other ensembles are not expected to be symmetric. The excited-state contaminations are slightly larger on the 32ID and 32If ensembles since the effective smearing sizes are smaller compared to those on the other ensembles.

\begin{figure}[htb]
\centering
\includegraphics[scale=0.7]{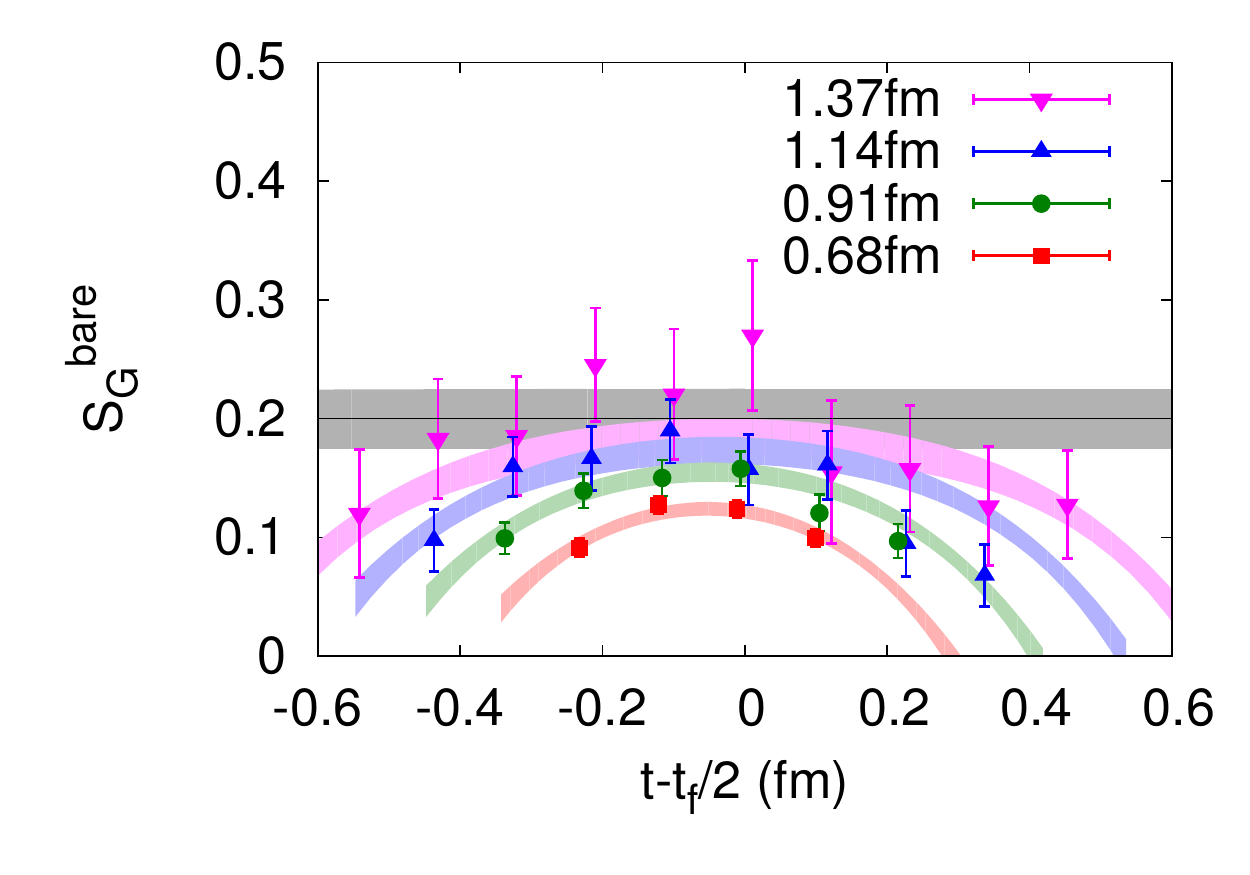}
\includegraphics[scale=0.7]{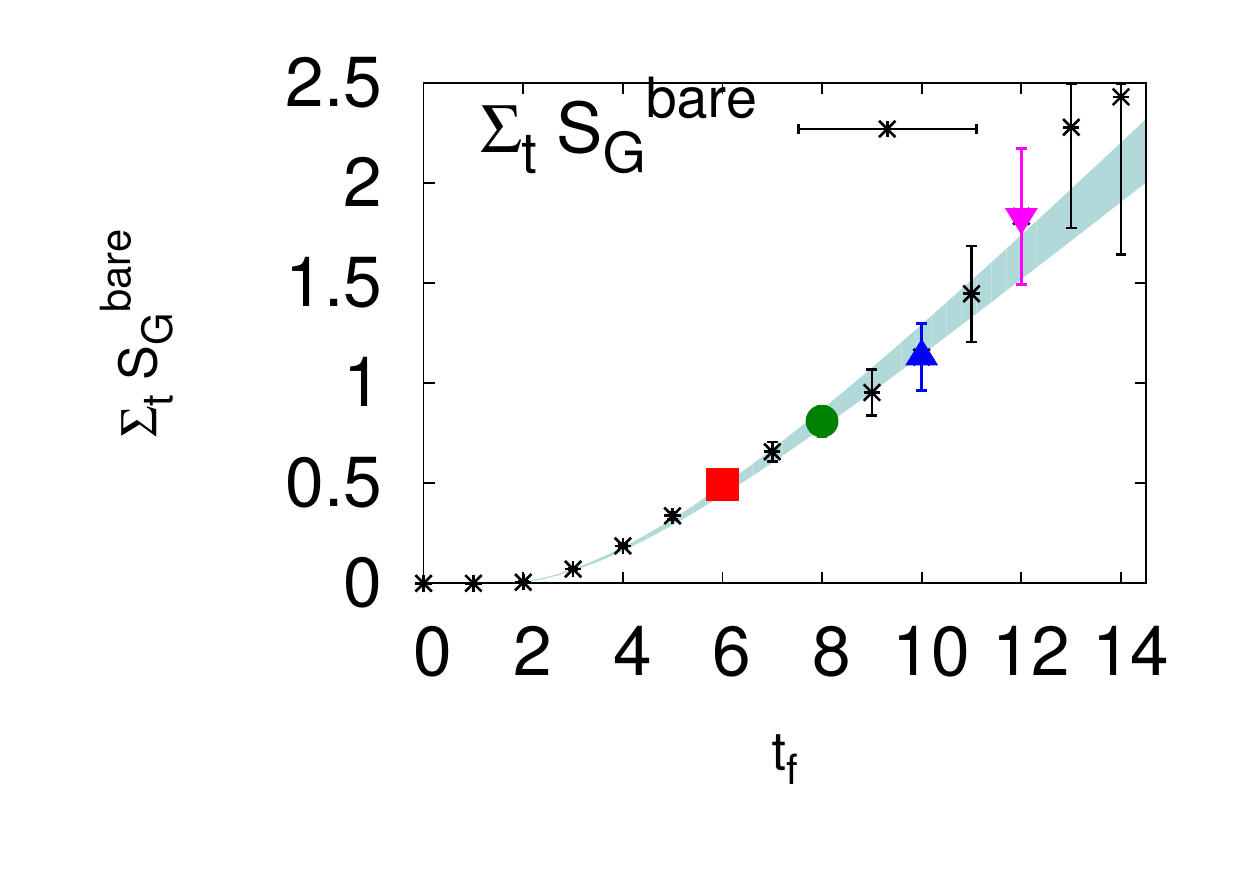}
\vspace*{-0.5cm}
 \caption{\small The summed ratio together with the plot of the ratios themselves for the case of $m_{\pi}=321$ MeV on the 24I ensemble. The left panel shows the ratio $R(t_f,t)$ as a function of the source-sink separation $t_f$ and the current time slice $t$, for the bare glue spin matrix element in the proton, $S_{G}^{bare}$. The gray band in the left panel  shows the result extrapolated to infinite separation, which corresponds to the prediction of $S_{G}^{bare}$. In the right panel, the four colored points correspond to the sum of the ratio at four different values of $t_f$ shown in the left panel and the band shows the prediction of $SR(t_f)\equiv\sum_t S_{G}^{bare}(t_f,t)$ based on the two-state fit in Eq.~\ref{eq:two_state2}. The data points agree with the band very well at larger $t_f$.}
\label{fig:24I}
\end{figure}

\begin{figure}[htb]
\centering
\includegraphics[scale=0.7]{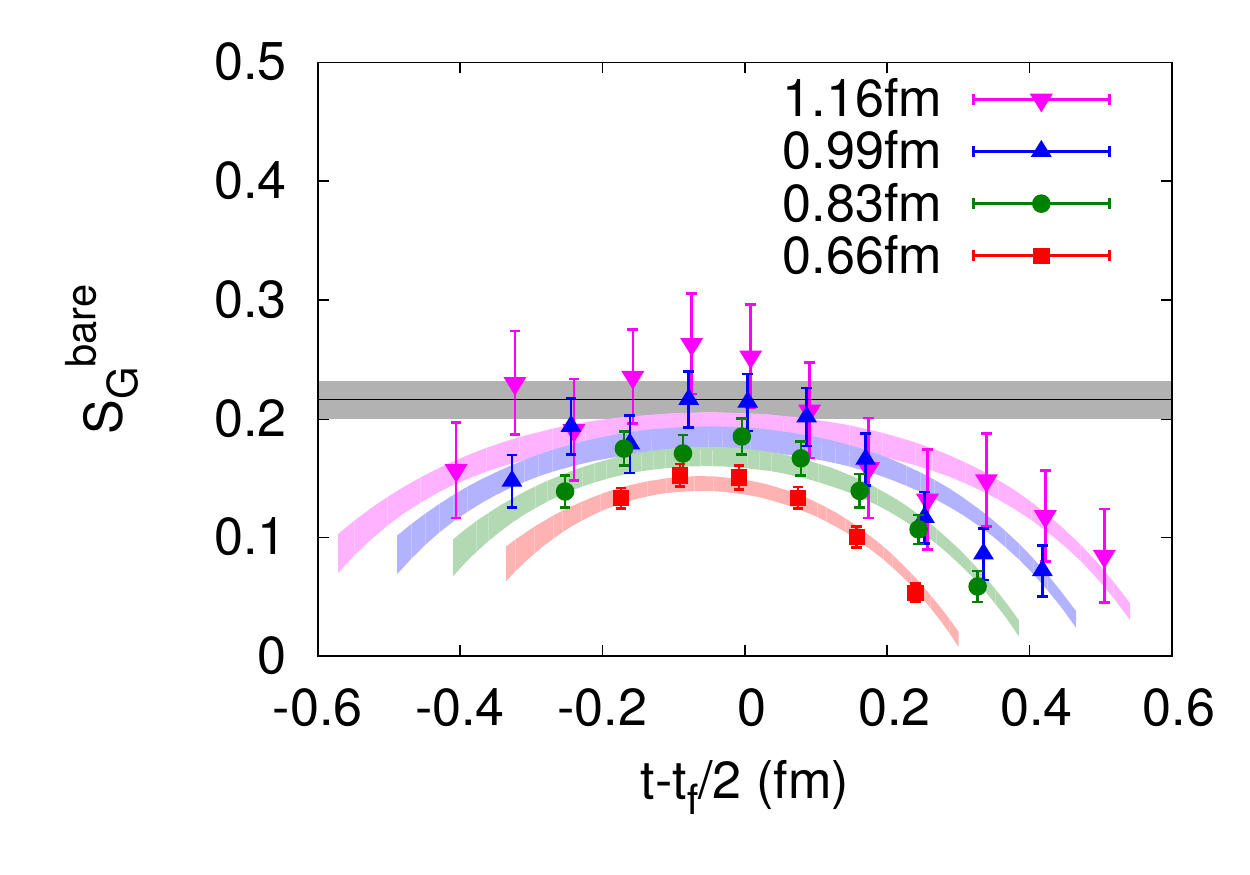}
\includegraphics[scale=0.7]{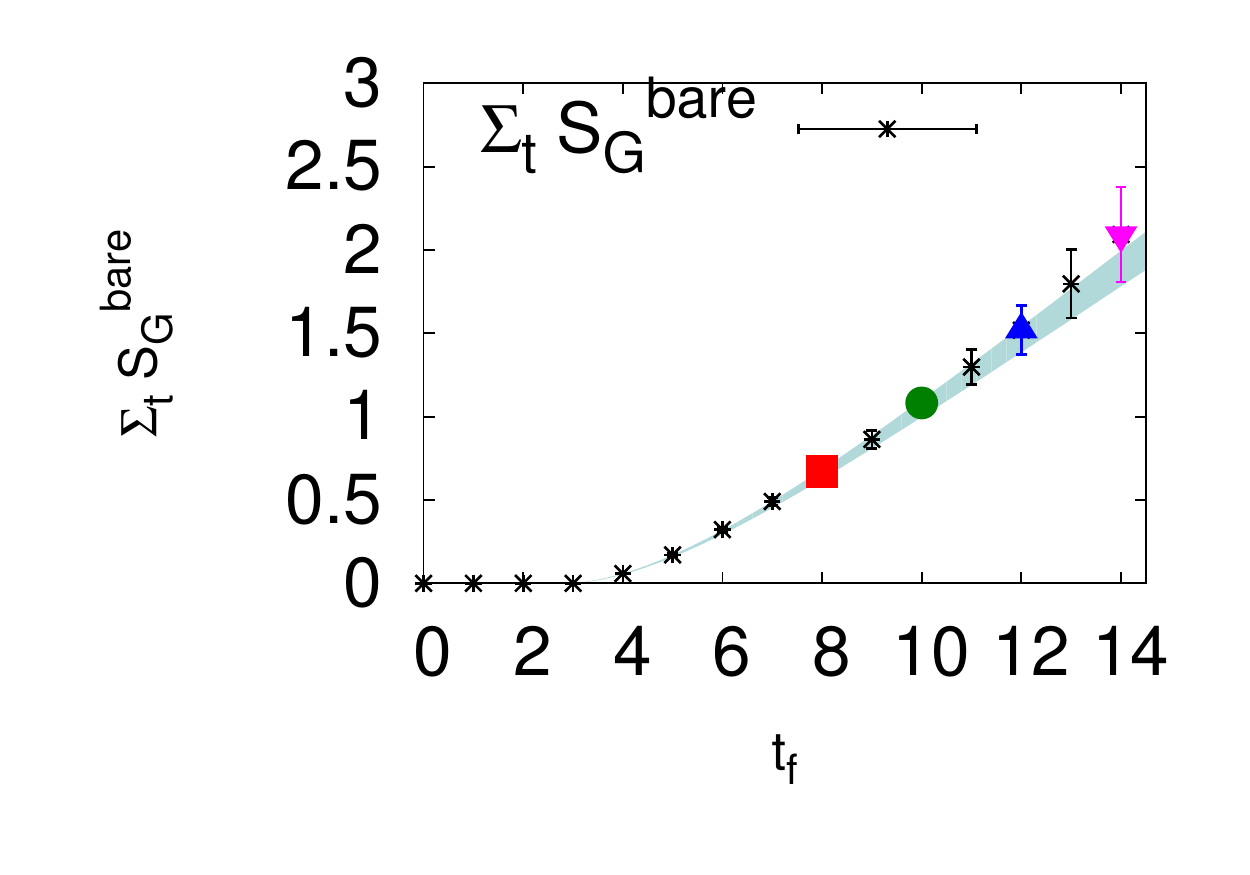}
\vspace*{-0.5cm}
 \caption{\small The same plot for the case of $m_{\pi}=315$ MeV on the 32I ensemble.}
\label{fig:32I}
\end{figure}

\begin{figure}[htb]
\centering
\includegraphics[scale=0.7]{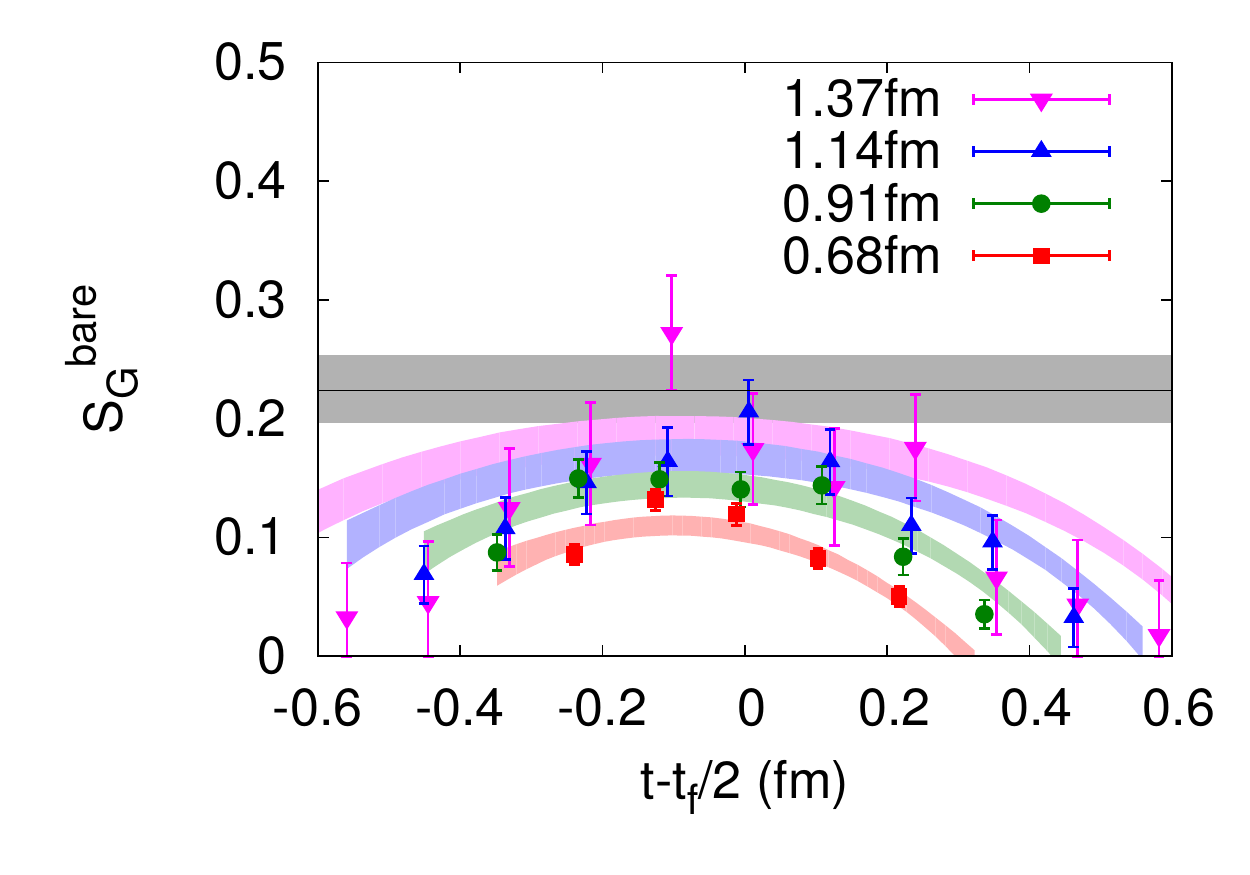}
\includegraphics[scale=0.7]{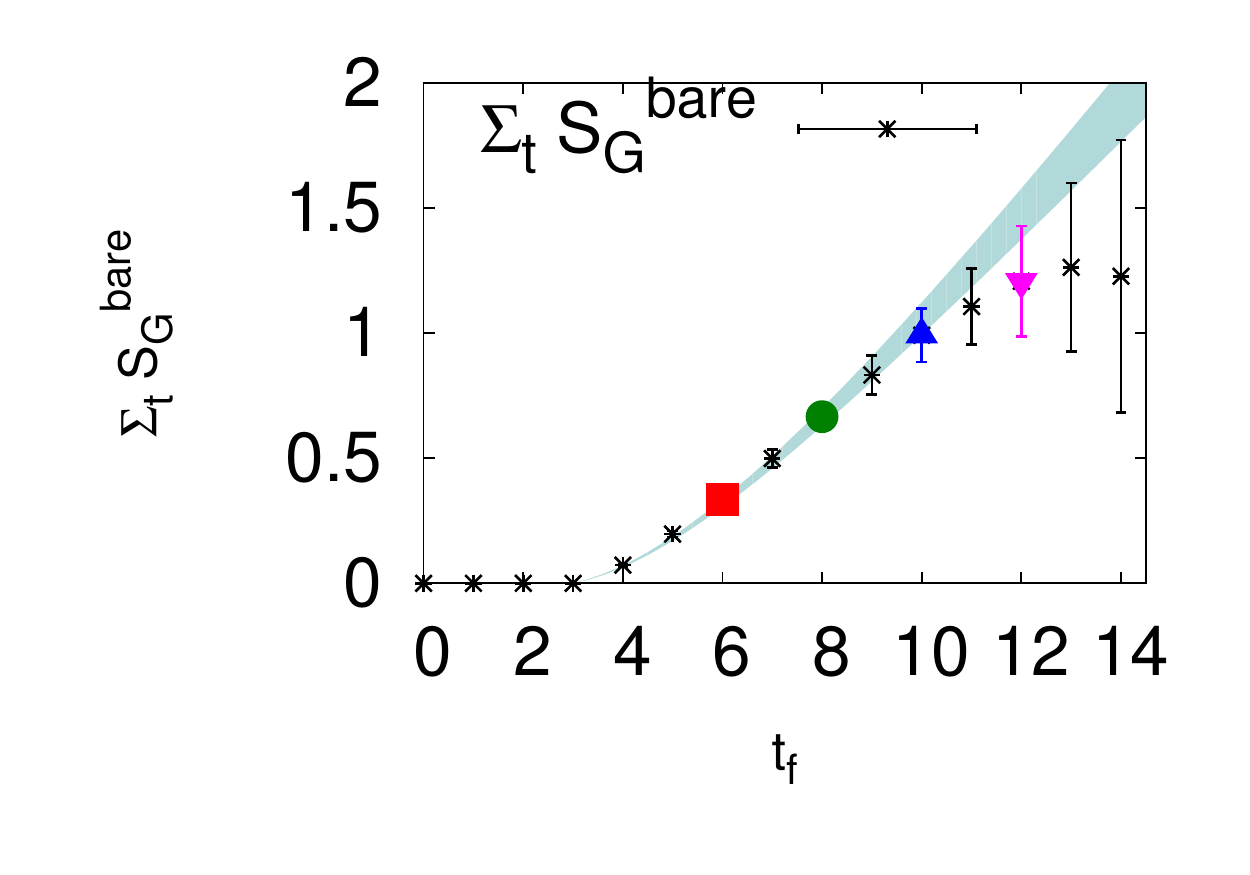}
\vspace*{-0.5cm}
 \caption{\small The same plot for the case of $m_{\pi}=326$ MeV on the 48I ensemble.}
\label{fig:48I}
\end{figure}

\begin{figure}[htb]
\centering
\includegraphics[scale=0.7]{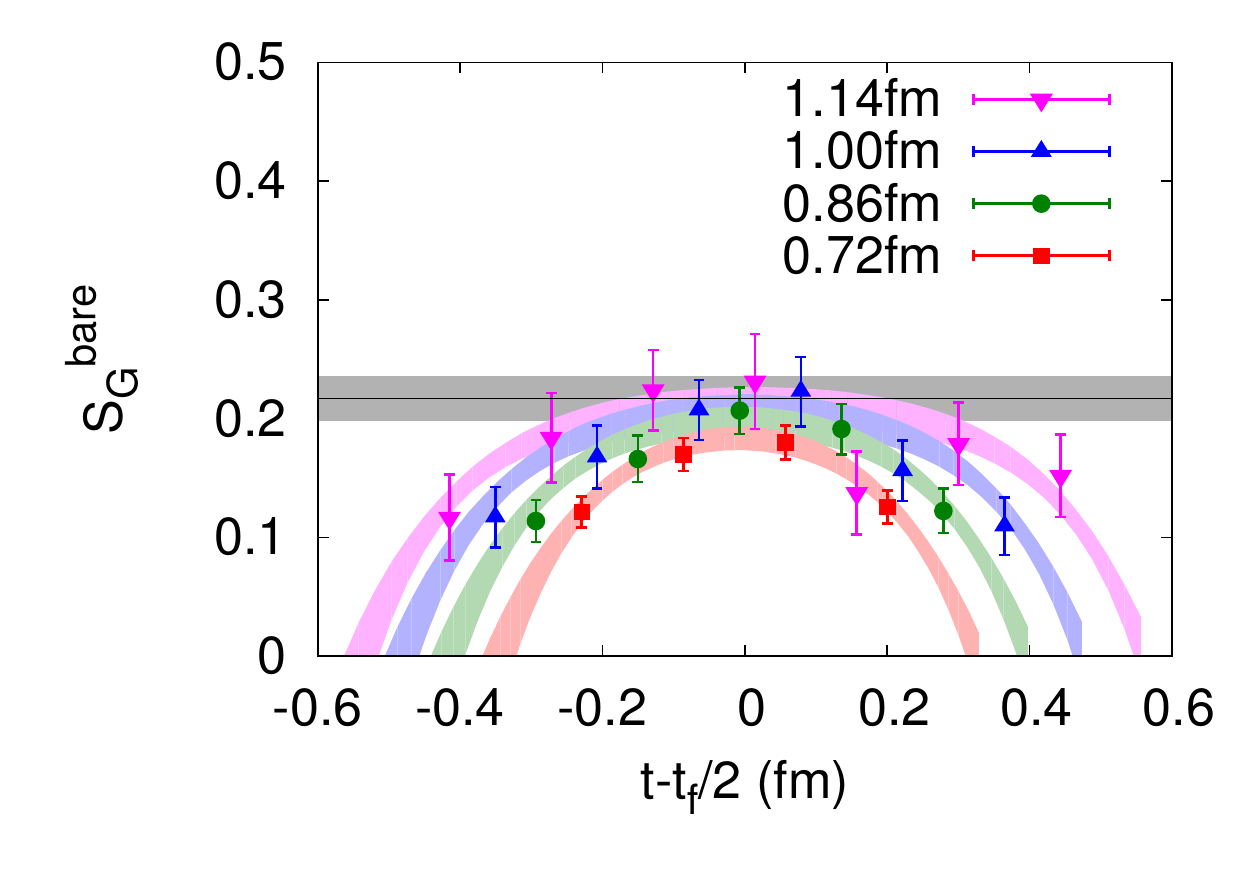}
\includegraphics[scale=0.7]{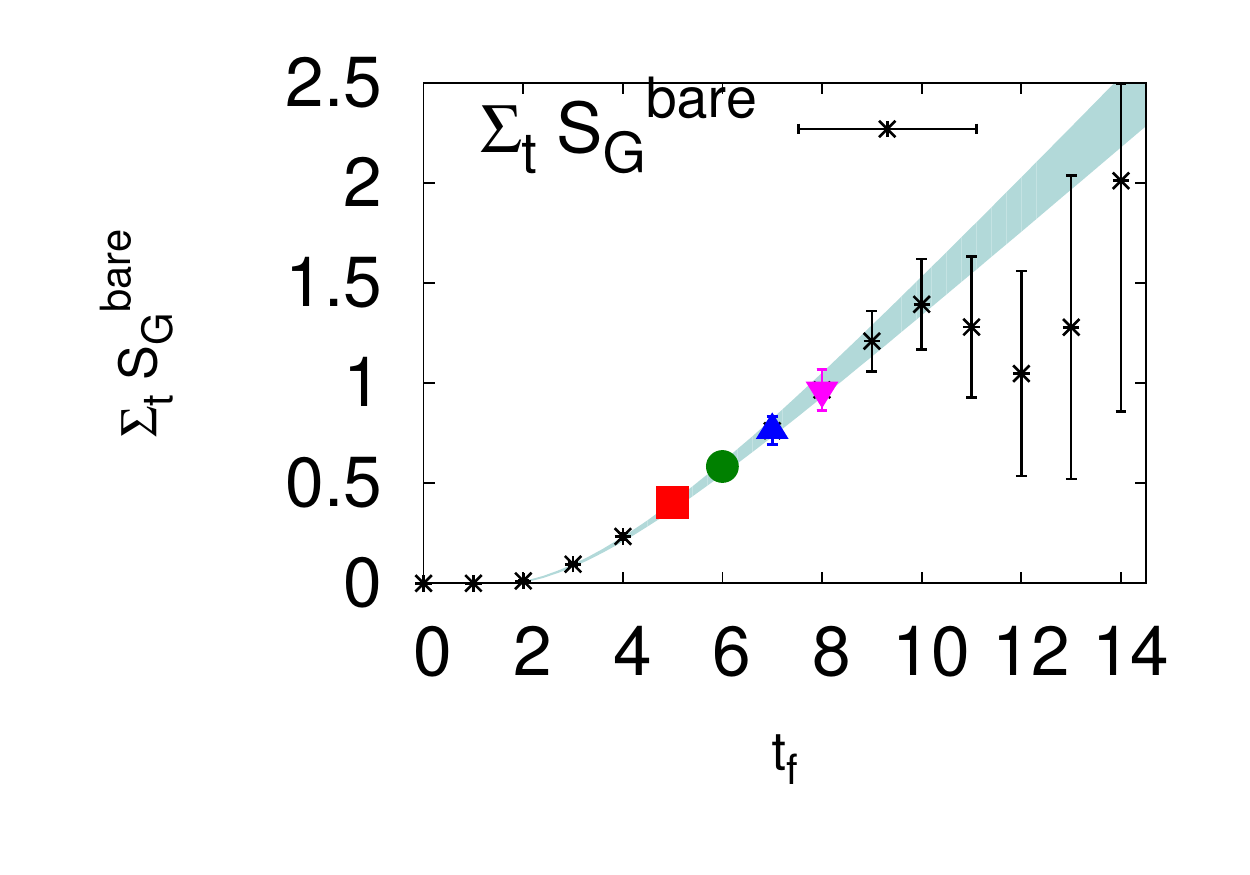}
\vspace*{-0.5cm}
 \caption{\small The same plot for the case of $m_{\pi}=331$ MeV on the 32ID ensemble.}
\label{fig:32ID}
\end{figure}

\begin{figure}[htb]
\centering
\includegraphics[scale=0.7]{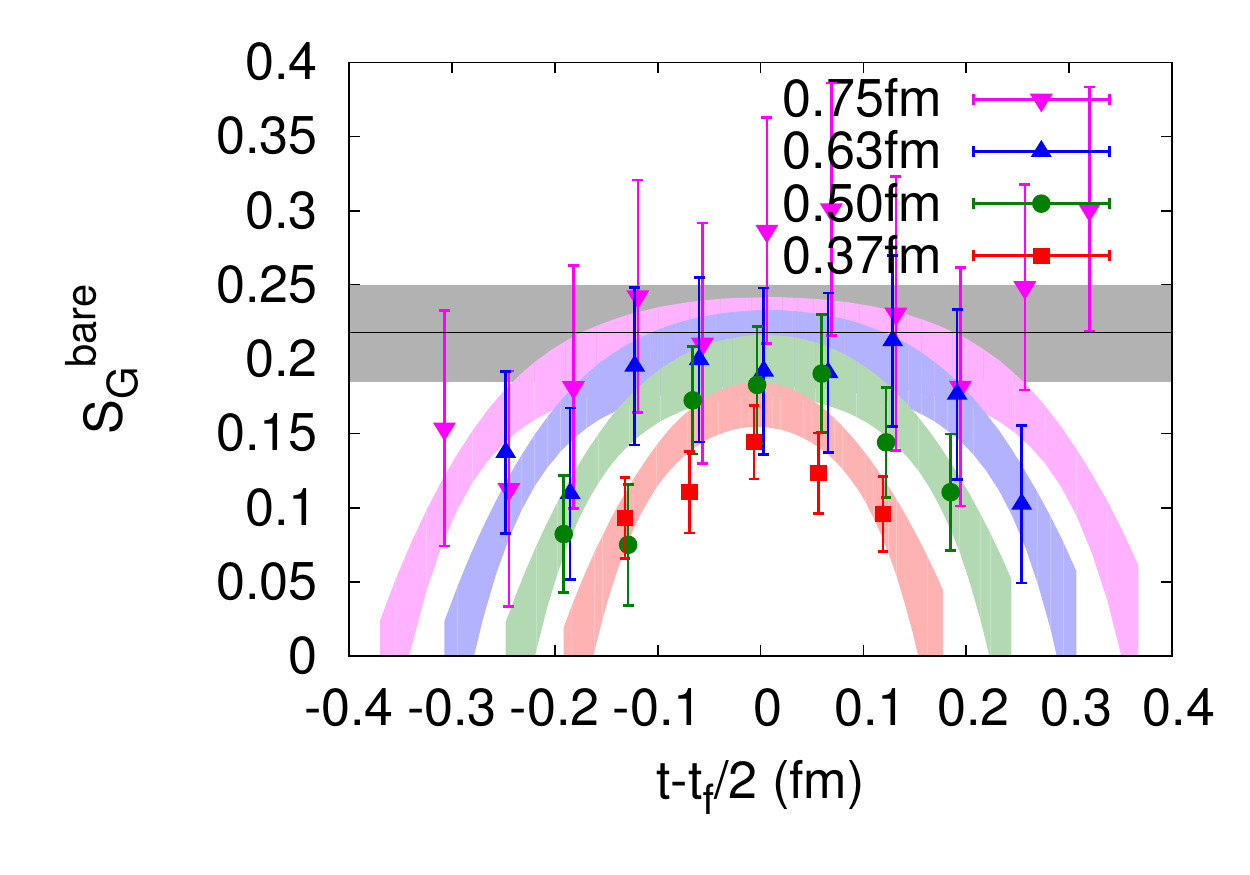}
\includegraphics[scale=0.7]{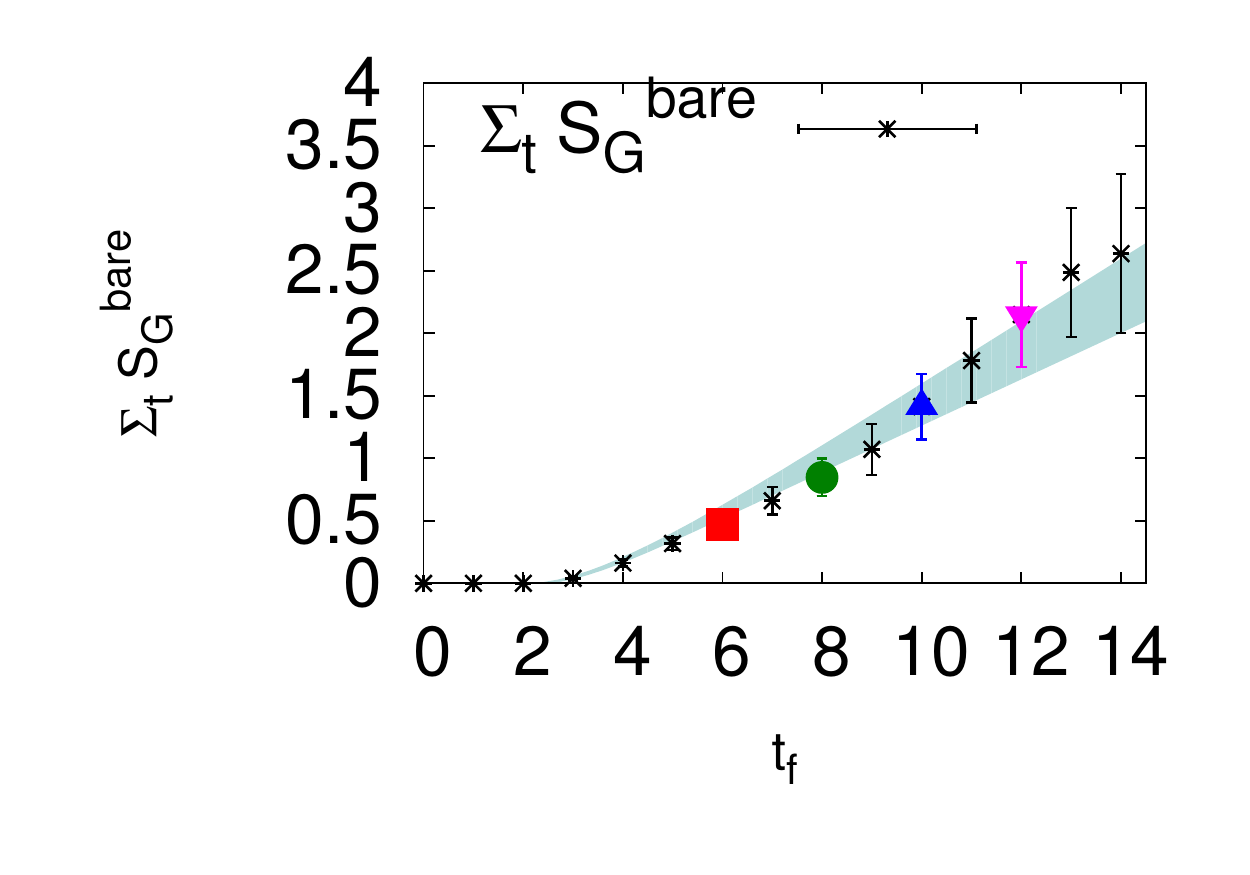}
\vspace*{-0.5cm}
 \caption{\small The same plot for the case of $m_{\pi}=340$ MeV on the 32If ensemble.}
\label{fig:32If}
\end{figure}

\end{widetext}
\clearpage

\bibliographystyle{apsrev4-1}
\bibliography{reference.bib}

\end{document}